\documentclass[pra,twocolumn,showpacs,superscriptaddress,
groupeaddress, preprintnumbers,amsmath,amssymb,tightenlines]{revtex4}
\usepackage{graphicx}
\newcommand{\beq}{\begin{equation}}
\newcommand{\eeq}{\end{equation}}
\newcommand{\bea}{\begin{eqnarray}}
\newcommand{\eea}{\end{eqnarray}}
\newcommand{\ba}{\begin{array}}
\newcommand{\ea}{\end{array}}
\newcommand{\bc}{\begin{center}}
\newcommand{\ec}{\end{center}}

\newcommand{\bml}{\begin{subequations}}
\newcommand{\eml}{\end{subequations}}
\newcommand{\commentout}[1]{{}}
\newcommand{\bk}{{\bf k}}

\newcommand{\br}{{\bf r}}

\newcommand{\K}{{\cal K}}
\newcommand{\adag}{a^\dagger}

\newcommand{\psidag}{\psi^\dagger}

\newcommand{\half}{\hbox{$\frac{1}{2}$}}

\newcommand{\HC}{{\rm H.c.}}
\newcommand{\HD}{{\cal H}}
\newcommand{\eq}[1]{(\ref{#1})}
\newcommand{\etal} {{\it et al.\/}}

\newcommand{\vol}[1]{{\bf #1}}
\newcommand{\comment}[1]{{}}
\newcommand{\rb}{$^{87}$Rb }
\newcommand{\pot}{$^{40}$K }

\begin{document}
\title{Raman Photoassociation of Bose-Fermi Mixtures and the Subsequent
Prospects for Atom-Molecule Cooper Pairing}

\author{Matt Mackie}
\affiliation{Department of Physics, University of Turku, FIN-20014
Turun yliopisto, Finland}
\affiliation{Helsinki Institute of Physics, PL 64, FIN-00014
Helsingin yliopisto, Finland}
\author{Olavi Dannenberg}
\affiliation{Helsinki Institute of Physics, PL 64, FIN-00014
Helsingin yliopisto, Finland}
\author{Jyrki Piilo}
\altaffiliation{Presently at School of Pure and Applied Physics,
University of KwaZulu-Natal, Durban 4041, South Africa.}
\author{Kalle-Antti Suominen}
\affiliation{Department of Physics, University of Turku, FIN-20014
Turun yliopisto, Finland}
\affiliation{Helsinki Institute of Physics, PL 64, FIN-00014
Helsingin yliopisto, Finland}
\author{Juha Javanainen}
\affiliation{Department of Physics, University of Connecticut,
Storrs, Connecticut, 06269-3046}
\date{\today}

\begin{abstract}
We theoretically investigate Raman photoassociation of a degenerate
Bose-Fermi mixture of atoms and the subsequent prospect for anomalous
(Cooper) pairing between atoms and molecules. Stable
fermionic molecules are created via free-bound-bound
stimulated Raman adiabatic passage which, in contrast to
purely bosonic systems, can occur in spite of collisions.
With the leftover atomic condensate to enhance intrafermion
interactions, the superfluid transition to atom-molecule Cooper pairs
occurs at a temperature that is roughly an order of magnitude below what
is currently feasible.
\end{abstract}


\maketitle

Photoassociation occurs when two atoms absorb a laser photon~\cite{WEI99},
thereby jumping from the free two-atom continuum to a bound molecular
state. Neutral-atom statistics is determined
by the number of neutrons in the nucleus--odd for fermions
and even for bosons. Similarly, the sum of the total number
of neutrons in the nuclei of the constituent atoms determines
neutral-molecule statistics. Molecules formed by
photoassociation of two fermions will accordingly result in a
boson, whereas fermionic molecules are born of a boson and
a fermion. Given degenerate
Bose-Fermi atoms, two questions
arise:  Will the atoms photoassociate with
into an arbitrary number of stable Fermi molecules? If so, is
it possible to realize atom-molecule Cooper pairing?

First introduced to explain
superconductivity, anomalous quantum correlations between
two degenerate electrons with equal and opposite momenta--Cooper
pairs--are due physically to an electron-electron attraction
mediated by the exchange of lattice-vibration-generated
phonons~\cite{BAR57}, and are a manifestation of fermionic
superfluidity~\cite{TIN75}. Anomalous pairing between different chemical
species was immediately suggested to explain the larger excitation energy
for nuclei with even rather than odd numbers of
nucleons~\cite{BOH58}, although it turned out that interspecies
pairing plays the dominant role.
Today quantum matter optics offers a means to explore condensed-matter
and nuclear physics by proxy, such as the pairing of fermions in atomic
traps and nuclei~\cite{HEI03}.

Here we investigate Raman photoassociation~\cite{VAR97,MAC00,HOP01,DAM03}
of Bose-Fermi mixtures of atoms~\cite{TRU01}, and the subsequent
prospects for Cooper pairing between different chemical species (i.e.,
atoms and molecules). First, we demonstrate that an arbitrary number of
stable Fermi molecules can be created via fractional~\cite{MAR91}
stimulated Raman adiabatic passage (STIRAP~\cite{BER98}), which is
feasible because, contrary to bosonic systems~\cite{HOP01}, collisional
interactions can be negligible. Density fluctuations in the
condensate leftover from the photoassociation process then replace the
vibrating ion lattice of the superconductor~\cite{HEI00b}, and the
subsequent phonon exchange can enhance the intrafermion attraction. We
find that a typical attraction is enhanced, but this enhancement is
insufficient for a transition to atom-molecule Cooper pairs within reach
of present ultracold technology.

We model
a Bose-Fermi mixture of atoms
coupled by heteronuclear photoassociation to electronically-excited
Fermi molecules, which is favored over homonuclear transitions
for well resolved resonances~\cite{SCH02}. The excited molecules
are themselves coupled by a second laser to electronically stable
molecules. For a degenerate system, the bosonic [fermionic] atoms
of mass $m_0$ [$m_+$] are represented by the field
$\psi_0(\br,t)$ [$\psi_+(\br,t)$], and the excited [stable]
fermionic molecules of mass $m_e=m_0+m_+$ [$m_-=m_e$] by the field
$\psi_e(\br,t)$ [$\psi_-(\br,t)$], with the boson (fermion) field operator
obeying commutation (anticommutation) relations. 
The Hamiltonian density for said non-ideal system is $\HD=\HD_0+\HD_I$,
where
\bml
\bea
\frac{\HD_0}{\hbar} &=& -\Delta\psidag_-\psi_-
  +(\delta-\Delta)\psidag_e\psi_e
   +\lambda_{+-}\psidag_+\psidag_-\psi_-\psi_+
\nonumber\\
  &&+\sum_\sigma\psidag_\sigma
    \left[-\frac{\hbar\nabla^2}{m_\sigma}-\mu_\sigma
   +\lambda_{0\sigma}\psidag_0\psi_0\right]\psi_\sigma\,,
\label{H0}
\\
\frac{\HD_I}{\hbar} &=& -\half\left[\left(\K_+\psidag_e\psi_+\psi_0
  +\Omega_-\psi_e\psidag_-\right) + \HC\right].
\eea
\label{FULL_HD}
\eml

The light-matter coupling due to
laser 1~(2) is ${\cal K}_+$ ($\Omega_-$), and the intermediate
(two-photon) laser detuning, basically the binding energy of the excited
(stable) molecular state relative to the photodissociation threshold, is
$\delta$ ($\Delta$). Particle trapping is implicit to the chemical
potential $\hbar\mu_\sigma$ ($\sigma=0,e,\pm$), and explicit traps can be
neglected for most practical purposes. Low-energy ($s$-wave) collisions
are accounted for by the boson-boson (boson-fermion, fermion-fermion)
interaction strength
$\lambda_{00}=2\pi\hbar a_{00}/m_0$
($\lambda_{0\pm}=4\pi\hbar a_{0\pm}/m_{0\pm}^*$,
$\lambda_{+-}=4\pi\hbar a_{+-}/m_{+-}^*$),
with $a_{\sigma_1\sigma_2}$ the $s$-wave scattering length
and $m_{\sigma_1\sigma_2}^*$ the reduced mass. Spontaneous decay,
included as $\Im[\delta]=-\Gamma$, is generally large enough
to justify the exclusion of excited-molecule collisions.

Consider now the process of stimulated Raman
adiabatic passage from atoms to stable
molecules. The key to STIRAP is
{\em counterintuitive} pulse timing, meaning that
the two lasers are adjusted so that in the beginning, when the population
is in the initial state (atoms), the strongest coupling is between the
intermediate and final states (excited and stable molecules), while
in the end, when effectively everything is in the target state, the
coupling between the initial and excited states is
strongest. As the system passes from atoms to stable molecules, the
state with the larger population is always weakly
coupled to the electronically-excited molecular state, and
the subsequently
low (ideally zero) population reduces (eliminates)
radiative losses.

An overview of our STIRAP model is now presented (see also Appendix).
When the number of bosons is much greater than the number of fermions,
$N_B\gg N_F$, the frequency scale for atom-molecule conversion is set by
$\Omega_+=\sqrt{\rho_B}\K_+$. Although a maximum
$N_B=100$ is used, qualitative scaling to large boson number is cinched
by assuming a density, $\rho_B=5\times 10^{14}\,{\rm cm}^{-3}$,
consistent with
$N_B=1.3\times 10^6$ Bose-condensed \rb atoms in a trap with radial and
axial frequencies
$\omega_r=100\times 2\pi\,$Hz and
$\omega_a=10\times 2\pi\,$Hz. For this density, a ballpark peak value for
the atom-molecule coupling is $\Omega_+\sim\Omega_0=2\pi\,$MHz.
A typical spontaneous decay rate is $\Gamma=10\times 2\pi\,$MHz.
The (assumedly negative) Fermi atom-molecule scattering length is
estimated as $|a_{+-}|=a_{00}=5.29\,$nm. The number of fermions is
restricted to $N_F=4$ for numerical ease, and large-particle-number
scaling is again ensured by picking a density,
$\rho_F=1.1\times 10^{12}\,\rm{cm}^{-3}$, consistent with $N_F=5\times
10^3$ \pot atoms in the same (mass-adjusted) trap as the bosons, so that
$\Lambda_{+-}=\lambda_{+-}/V=5.81\times 2\pi\,$Hz. Numerics are further
eased by restricting collisions between fermions to a bare minimum:
$\bk_1+\bk_2=\bk_3+\bk_4$; $\bk_1=\bk_3$,
$\bk_2=\bk_4$. Also, $N_B\gg N_F$ means that collisions with condensate
atoms can be neglected.

\begin{figure}
\centering
\includegraphics[width=8.0cm]{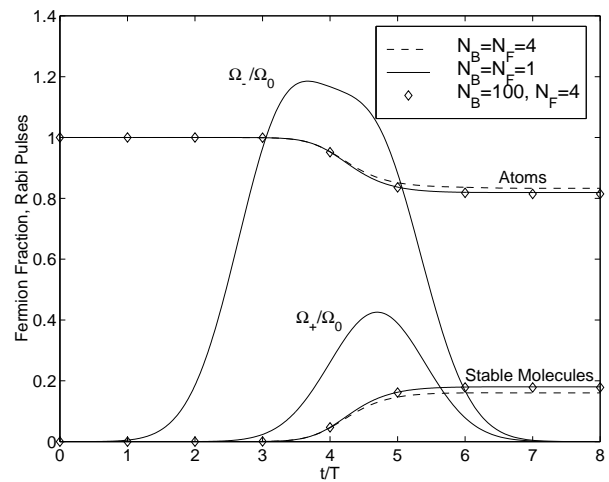}
\caption{
Creation of an arbitrary mixture of Fermi
degenerate atoms and molecules via fractional stimulated Raman adiabatic
passage.  Units of frequency are set by the choice
$\Omega_0=2\pi\,\rm{MHz}=1$, and
the pulse parameters are $\alpha=0.14\pi$, $T=5\times 10^3$, and
$\tau=0.7T$. For $N_B=100$ and $N_F=4$, fractional STIRAP exhibits no
visible difference from the $N_B=N_F=1$ case, while $N_B=N_F=4$
illustrates that many-body effects can limit the conversion
efficiency.}
\label{STIRAP}
\end{figure}

Of course, complete conversion would leave no
atoms to form Cooper pairs with molecules, so we pursue
fractional STIRAP~\cite{MAR91} via the Vitanov~\etal~\cite{MAR91}
pulseforms:
$\Omega_+(t)=\Omega_0\sin\alpha\exp[-\left(t-\tau\right)^2/T^2]$ and
$\Omega_-(t)=\Omega_0\exp[-\left(t+\tau\right)^2/T^2]
  +\Omega_0\cos\alpha\exp[-\left(t-\tau\right)^2/T^2]$, where
$\tan\alpha$ sets the final population fraction. Using Fock states and
the Hamiltonian given in the Appendix, we solve the Schr\"odinger
equation numerically. The key results are presented in Fig.~\ref{STIRAP}.
For $N_F=4$ and $N_B=100$, the system reproduces the single particle case
($N_B=N_F=1$), i.e., the results are identical to those for a three-level
atom, as expected for a mostly-undepleted boson field. Atom-molecule
collisions are negligible for couplings as large as
$\Lambda_{+-}/\Omega_0=10^{-5}$, which is similar to a bosonic system,
lending confidence to the restricted-collision model.  Furthermore,
$N_F=N_B=4$ shows that many-body effects can limit molecular conversion.
This many-body effect is similar to the one-color BEC
case~\cite{JAV99}, and is not attributable to fermion statistics.

It is not entirely surprising to find a fermionic atoms responding
as a unit to form fermionic molecules: cooperative fermionic
behavior was predicted for one-color formation of Fermi
molecules~\cite{DAN03}, as well as in atomic four-wave
mixing~\cite{MOO01} and, more relatedly, cooperation was observed in
magnetoassociation of Fermi atoms into Bose molecules~\cite{REG03}.
However, what is surprising is that STIRAP should work basically as
expected {\em even in the presence of collisions}. This
situation arises because, when $N_B\gg N_F$, the condensate density is
effectively fixed, and the associated mean-field contribution simply
amounts to a static bias that can be absorbed into the detuning; with
Fermi-Fermi collisions blocked, which is most likely for small
ground-state molecules, only collisions between the Fermi atoms and
molecules can move the system off the required two-photon resonance, and
STIRAP works better. In other words, we get the advantage of Bose
enhancement of the free-bound coupling ($\Omega_+\propto\sqrt{\rho_B}$),
without the mean-field shifts. While limited computational resources
preclude explicit investigation, these results ought to scale
qualitatively with increasing particle number, as we have seen for up to
$N_F=20$ in one-color production of Fermi molecules.

Now we are safe to presume the existence of an arbitrary admixture
of Fermi-degenerate atoms and molecules, and thus to consider any
subsequent anomalous pairing. Once the transient STIRAP pulses have
vanished, the system is described by $\HD_0$ [Eq.~\eq{H0}] with
$\Delta=\delta=0$ and
$\sigma=0,\pm$. For equal-mass fermions, it is known that a fermion
density fluctuation gives rise to an effective chemical potential for the
bosons, which creates a bosonic density fluctuation, which in turn leads
to an effective chemical potential for the fermions. In other words,
phonons spawned by BEC density fluctuations are exchanged between the
fermions, altering their interaction. Just like lattice vibrations that
drive the attraction between degenerate electrons in superconductors, BEC
density fluctuations lead to an attractive interaction that can enhance
overall attractions, and thus Cooper pair formation~\cite{HEI00b}. 

Here the effective Fermi-Fermi scattering length is
\beq
\bar{a}_{+-} = a_{+-}
  \left[ 1+\frac{\ln(4e)^{2/3}}{\pi}\,k_Fa_{+-}
 -H\,\frac{\lambda_{0+}\lambda_{0-}}{\lambda_{00}\lambda_{+-}}\right],
\label{A_EFF}
\eeq
where
$H=\ln(1+x^2)/x^2$ with $x=\hbar k_F/m_0v_s$ and
$v_s=(\rho_B\hbar\lambda_{00}/m_0)^{1/2}$ is the speed of phonons
in BEC; $\hbar k_F\ll m_0v_s$ implies $H\approx 1$. In other words, the
effective scattering length can be written
$\bar{a}_{+-}/a_{+-}=1+\eta_{FF}-\eta_{FB}$, where $\eta_{FF}$
($\eta_{FB}$) is the contribution to atom-molecule interactions
from fermion-fermion (boson-fermion) fluctuations. Implicit to
expression~\eq{A_EFF} is the perturbative assumption
$\eta_{FB}\ll\eta_{FF}$. The immediate contrast with Ref.~\cite{HEI00b} is
that $\eta_{FB}<0$ is allowed.
For a weakly attractive system
($k_F|a_{\sigma_1\sigma_2}|,\rho_B|a_{\sigma_1\sigma_2}|^3\ll 1$), the
critical temperature for Cooper pairing is
\beq
T_c = 0.61\,T_F\,
  \exp\left[-\frac{\pi/4}{k_F|\bar{a}_{+-}|}\right],
\label{BCS_EQ}
\eeq
where
$T_F=\hbar(\mu_++\mu_-)R_M/k_B$ is the Fermi temperature with
$R_M=m_{+-}^*/\sqrt{m_+m_-}$. The Fermi wavevector, $k_F$, was taken as
the same for both species, so that $\mu_++\mu_-=(m_\pm/m_{+-}^*)\mu_\pm$
and $T_F=T_F^{(+)}\sqrt{m_+/m_-}\,$.

Continuing to focus on $^{87}$Rb-$^{40}$K,
$N_B=1.3\times 10^6$ BEC
atoms in a trap with $\omega_r=100\times 2\pi\,$Hz and
$\omega_a=10\times 2\pi\,$Hz yield a boson density 
$\rho_B=5\times 10^{14}\,\rm{cm}^{-3}$. A modest number of fermions,
$N_F=5\times 10^3$, means that the loss of condensate atoms in
molecule formation can be neglected, the condensate will absorb any heat
created by pairing-induced holes in the atomic Fermi sea~\cite{TIM01b},
and collapse instabilities~(Modugno \etal~\cite{TRU01}) are avoided.
Presuming that fractional STIRAP converts roughly 18\% of the initial
Fermi atoms into molecules (see Fig.~1), and that the fermions share
the same (mass-adjusted) trap, then
$\rho_\pm=1.1\times 10^{12}\,{\rm cm}^{-3}$, and the requirement of
equal Fermi wavenumbers for the atoms and molecules is met. For the
given parameters, the size of the BEC is
roughly an order of magnitude larger than the Fermi clouds, so that
overlap should not be an issue.

\begin{figure}
\centering
\includegraphics[width=8.0cm]{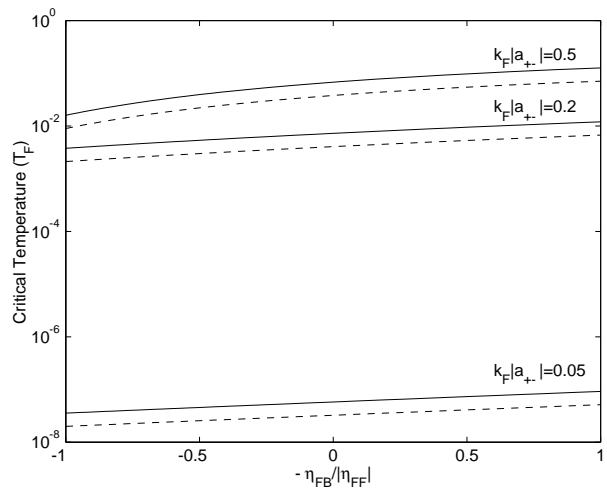}
\caption{Critical temperature for the superfluid transition to anomalous
atom-molecule pairs as a function of fermion-boson fluctuation
strength.  Calculations are for weak ($k_F|a_{+-}|=0.05$), marginally weak
(0.2), and marginally strong (0.5) interactions. The solid (dashed) curve
gives the critical temperature in units of the effective (atomic) Fermi
temperature.}
\label{CRIT_T}
\end{figure}

Figure~\ref{CRIT_T} summarizes our investigations. Under the above trap
conditions, and for $|a_{+-}|=a_{00}=5.29\,\rm{nm}$, we find the
weakness parameter $k_F|a_{+-}|=0.05$ and an unreasonably low
critical temperature. Nevertheless, if
the trap is modified to be anharmonic, then tighter confinement
ultimately means a diluteness parameter on the order of
$k_F|a_{+-}|=0.2$~\cite{SEA02}, and the situation is improved but still
out of reach of current technology
($T\sim0.05T_F$~\cite{REG03}). The only chance appears
to be for a tight trap and a large scattering length,
$k_F|a_{+-}|=0.5$; however, the theory is at best marginally
applicable in this regime and, besides, some other process (e.g.,
three-body recombination) would probably win out before superfluidity
could set-in to such a system.

In conclusion, collisions can arguably be made negligible in
photoassociation of Bose-Fermi mixtures of atoms, and fractional
stimulated Raman adiabatic passage scheme is therefore feasible for
creating Fermi degenerate molecules. On the other hand,
while possible in principle, accessible atom-molecule superfluidity means
that next-generation technology must shed {\em another} order of
magnitude in temperature from the first generation~\cite{DEM99}, or that a
system with an attractive {\em and} a strong-but-not-too-strong
interaction will be found. Our opinion is that it will be worthwhile to
look for other ways of driving atom-molecule Cooper pairs, such as a
photoassociation or Feshbach resonance.

{\em Appendix.}--To model STIRAP from Eq.~\eq{FULL_HD}, make a
time-dependent unitary transformation $U(t)=\Pi_\sigma U_\sigma(t)$, where
$U_{0(+)}(t)=\exp[-i\mu_{+(0)}\psidag_{+(0)}\psi_{+(0)}t]$,
$U_e(t)=\exp[i(\mu_++\mu_0)\psidag_e\psi_et]$,
$U_-(t)=\exp[-i(\mu_++\mu_0)\psidag_-\psi_-t$, whereby
$\mu_e\rightarrow \mu'_e=\mu_e-(\mu_-+\mu_0)$ and
$\mu_-\rightarrow \mu'_-=\mu_--(\mu_-+\mu_0)$; in turn, absorb
$\mu'_{e(-)}$ into the intermediate (two-photon)
detuning. Assume that atom-molecule conversion occurs on a timescale much
faster than the motion of the atoms in the trap, and thereby neglect
the kinetic energies and any explicit trapping potentials.
Take the Fermi energy to lie within the Wigner threshold regime,
so that the coupling
$\K_+$ is the same for all modes. Focus on the regime $N_B\gg N_F$, so
that the condensate is practically undepleted and
$\psi_0\approx\sqrt{\rho_B}$ can be absorbed into the atom-molecule
coupling strength; likewise, the BEC mean-field shifts,
$\lambda_{0\sigma}\psidag_0\psi_0\approx\rho_B\lambda_{0\sigma}$, are a
constant that can be absorbed into the detuning, and therefore of no
concern. Input two-photon and intermediate resonance by setting $\Delta=0$ and
$\delta=-\half i\Gamma$. Finally, switch to momentum space, so that
\bea
\frac{H}{\hbar}&=&
  -\half i\Gamma\sum_\bk\adag_{\bk e} a_{\bk e}
    +\Lambda_{+-}\sum_{\{\bk_i\}}
    \adag_{\bk_1+}\adag_{\bk_2-}a_{\bk_3-}a_{\bk_4+}
\nonumber\\&&
\hspace{-0.1 cm}
  -\half\sum_\bk\left[\left(\Omega_+\adag_{\bk e} a_{\bk+}
      +\Omega_-\adag_{\bk e} a_{\bk-}\right) + \HC\right]\!,
\nonumber
\label{SIMP_HAM}
\eea
where $\Omega_+=\sqrt{\rho_B}\K_+$ and
$\Lambda_{+-}=\lambda_{+-}/V$. Recover the few-boson results
by assuming that Bose stimulation favors
condensate modes, and making the
substitutions $\Omega_+\rightarrow\Omega_+/\sqrt{N_B}$ and
$\adag_{\bk e} a_{\bk+}\rightarrow\adag_{\bk e} a_{\bk+}a_0$.

{\it Acknowledgements.--}
\noindent The authors thank Emil Lundh, Chris
Pethick, Mirta Rodriguez, Eddy Timmermans, and P\"aivi T\"orm\"a for
helpful discussions; also, the
Academy of Finland (MM and KAS; project 50314), the Magnus Ehrnrooth
foundation (OD), NSF and NASA (JJ; PHY-0097974 and NAG8-1428) for
financial support.


\begin{thebibliography}{99}
\vspace{0.075cm}

\bibitem{WEI99}
  J. Weiner, V. S. Bagnato, S. Zilio, and P. S. Julienne,
  {\rmp} {\bf 71}, 1 (1999).

\bibitem{BAR57}
J. Bardeen, L. N. Cooper, and J. R. Schrieffer,
   Phys. Rev. \vol{106}, 162 (1957).

\bibitem{TIN75}
  M.~Tinkham, {\it Intro. to Superconductivity},
    (McGraw-Hill, New York, 1975).

\bibitem{BOH58}
  A. Bohr, B. R. Mottelson, and D. Pines,
      {Phys. Rev.} \vol{110}, 936 (1958).

\bibitem{HEI03}
  H. Heiselberg, physics/0304005.

\bibitem{VAR97}
A. Vardi, D. Abrashkevich, E. Frishman, and M. Shapiro,
  \jcp {\bf 107}, 6166 (1997).

\bibitem{MAC00}
  M. Mackie, R. Kowalski, and J. Javanainen,
      {\prl} \vol{84}, 3803 (2000).

\bibitem{HOP01}
  J. J. Hope, M. K. Olsen, and L. I. Plimak,
    {\pra} {\bf 63}, 043603 (2001);
  P. D. Drummond, K. V. Kheruntsyan, D. J. Heinzen, and R. H. Wynar,
        \pra {\bf 65}, 063619 (2002);
  M. Mackie, A. Collin, and J. Javanainen, physics/0212111.

\bibitem{DAM03}
  B. Damski, L. Santos, E. Tiemann, M. Lewenstein,
    S. Kotochigova, P. Julienne, and P. Zoller,
    {\prl} \vol{90}, 110401 (2003).

\bibitem{TRU01}
  A. G. Truscott, K. E. Strecker, W. I. McAlexander, G. B. Partridge, and
    R. G. Hulet, Science {\bf 291}, 2570 (2001);
  F. Schreck, G. Ferrari, K. L. Corwin, J. Cubizolles, L. Khaykovich,
    M.-O. Mewes, and C. Salomon,
    {\pra} {\bf 64}, 011402 (R) (2001);
  G. Modugno, G. Roati, F. Riboli, F. Ferlaino, R. J. Brecha,
    and M. Inguscio, {Science} {\bf 287}, 2240 (2002).

\bibitem{MAR91}
  P. Marte, P. Zoller, and J. L. Hall,
    \pra\vol{44}, R4118 (1991);
  N. V. Vitanov, K.-A. Suominen, and B. W. Shore,
    {J.~Phys. B} \vol{32}, 4535 (1999).

\bibitem{BER98}
  K.~Bergmann, H. Theuer, and B. W. Shore,
  {\rmp} {\bf 70}, 1003 (1998).

\bibitem{HEI00b}
  H. Heiselberg, C. J. Pethick, H. Smith, and L. Viverit,
    {\prl} \vol{85}, 2418 (2000);
  M. J. Bijlsma, B. A. Heringa, and H. T. C. Stoof,
    {\pra} \vol{61}, 053601 (2000).

\bibitem{SCH02}
  U. Schl\"oder, C. Silber, T. Deuschle, and C. Zimmermann,
  \pra \vol{66}, 061403 (R) (2002).

\bibitem{JAV99}
  J. Javanainen and M. Mackie, \pra {\bf 59}, R3186 (1999);
A. Vardi, V. A. Yurovsky, and J. R. Anglin, \pra {\bf 64}, 063611 (2001).

\bibitem{DAN03}
  O. Dannenberg, M. Mackie, and K.-A. Suominen,
    \prl\vol{91}, 210404 (2003);
  M. Wouters, J. Tempere, and J. T. Devreese,
    \pra\vol{67}, 063609 (2003).

\bibitem{MOO01}
  M. G. Moore, and P. Meystre,
  {\prl} {\bf 86}, 4199 (2001);
  W. Ketterle, and S. Inouye,
  \prl\vol{86}, 4203 (2001).

\bibitem{REG03}
  C. A. Regal, C. Ticknor, J. L. Bohn, and D. S. Jin,
    Nature \vol{424}, 47 (2003);
  K. E. Strecker, G. B. Partridge, and R. G. Hulet,
    \prl\vol{91}, 080406 (2003).

\bibitem{TIM01b}
  E. Timmermans,
  {\prl} \vol{87}, 240403 (2001).

\bibitem{SEA02}
  C. P. Search, H. Pu, W. Zhang, B. P. Anderson, and P. Meystre,
    \pra\vol{65}, 063616 (2002).

\bibitem{DEM99}
  B. DeMarco and D. S. Jin, Science {\bf 285}, 1703 (1999).

\end{thebibliography}
\end{document}